\documentclass[aps,prb,twocolumn,superscriptaddress]{revtex4}
\usepackage{graphicx}
\usepackage{natbib}

\newcommand{\ket}[1]{| #1 \rangle}
\newcommand{\bra}[1]{\langle #1 |}
\newcommand{\rb}[1]{\left( #1 \right)}

\newcommand{\ex}[1]{\langle #1 \rangle}

\newcommand{\beq}{\begin{eqnarray}}
\newcommand{\eeq}{\end{eqnarray}}

\begin{document}

\title {Quantum chaos and critical behavior on a chip}
\author{Neill Lambert}
\affiliation{Advanced Science
        Institute,
     The Institute of Physical and Chemical Research (RIKEN), Saitama 351-0198, Japan}
\author{Yueh-nan Chen}
\affiliation{Department of Physics and National Center for
Theoretical Sciences, National Cheng-Kung University, Tainan 701,
Taiwan}
\author{Robert Johannsson}
\affiliation{Advanced Science
        Institute,
     The Institute of Physical and Chemical Research (RIKEN), Saitama 351-0198, Japan}
\author{Franco Nori}
\affiliation{Advanced Science
        Institute,
     The Institute of Physical and Chemical Research (RIKEN), Saitama 351-0198, Japan}
\affiliation{Center for Theoretical Physics, Physics Department,
Applied Physics Program, Center for the Study of Complex Systems,
The University of Michigan, Ann Arbor, Michigan, 48109-1040, USA}

\begin{abstract}
The Dicke model describes $N$ qubits (or two-level atoms)
homogenously coupled to a bosonic mode.  Here we examine an
open-system realization of the Dicke model, which contains critical
and chaotic behaviour. In particular, we extend this model to
include an additional open transport qubit (TQ) (coupled to the
bosonic mode) for passive and active measurements. We illustrate how
the scaling (in the number of qubits $N$) of the superradiant phase
transition can be observed in both current and current-noise
measurements through the transport qubit.  Using a master equation,
we also investigate how the phase transition is affected by the
back-action from the transport qubit and losses in the cavity.  In
addition, we show that the non-integrable quantum chaotic character
of the Dicke model is retained in an open-system environment. We
propose how all of these effects could been seen in a circuit QED
system formed from an array of superconducting qubits, or an atom
chip, coupled to a quantized resonant cavity (e.g., a microwave
transmission line).
\end{abstract}

\maketitle

\section{Introduction}

Understanding and categorizing complex modes of behavior, such as
quantum phase transitions~\cite{Sachdev99} and quantum
chaos~\cite{Gutzwiller90}, is an important part of quantum many-body
theory.  
  Recently, concepts and formalisms from
quantum information theory have been used to understand and classify
several aspects of criticality \cite{Osterloh02,Osborne02,Vidal02,
Latorre03, Lambert04}. However, the realization of strong coupling
regimes, coherent dynamics, and careful readout necessary to observe
these phenomena in laboratory conditions is challenging.

Our goal here is to show how a particular quantum phase transition,
the Dicke superradiant transition \cite{HL73,Lambert04,Emary02}
behaves when coupled to the environment and measured using transport
techniques, as is the case in realistic experimental conditions. The
Dicke model describes $N$ two-level ``atoms'' or qubits coupled to a
common single-mode cavity.  We focus on this model because of the
recent advances in on-chip `circuit-QED'
\cite{You03,You032,you05,Blais2, Mooji, Takayanagi, Deppe}, where
the strong coupling regime is accessible, and which allow for
coupling to a range of artificial atoms and measurement apparatuses.
In particular, we propose a dispersive measurement scheme to observe
this transition by coupling either a superconducting qubit array, or
an atom chip, to a cavity which is simultaneously (dispersively)
coupled to a non-equilibrium measurement device (a so-called {\em
``transport'' qubit}~\cite{Lambert082}), realisable with a
superconducting single electron transistor, or double quantum dot.
The geometry of the proposed device is shown in Fig. 1 and described
in detail in its caption.

We begin by outlining the salient features of the phase transition
in the Dicke model, and existing work in this area (section II), and
discussing the closed (section III) and open (section IV)
descriptions of the model.  We then investigate how coupling to a
transport qubit allows readout of the phase transition properties,
and give analytical and numerical results for the current and
current-noise in the zero back-action limit (section V).  We then
discuss our main result: that the current can be used as an {\em
observable order parameter} to detect the phase transition (section
VI). This complements a recent surge of interest in identifying
signatures of complex behaviour in mesoscopic transport measurements
\cite{Lambert082, tobias04,blanter00}.  We then consider back-action
and decoherence (cavity-loss) effects using a master equation
approach (section VII). We show that both transport qubit
back-action and cavity loss appear to only have a weak affect on the
current measurement near the critical point. In addition, we show
that the Liouvillian describing the open-system dynamics has an
eigenvalue spectrum similar to that of the Wigner-Dyson distribution
of random matrix theory, as in the closed system case (section
VIII).
 Finally, we briefly discuss practical schemes to realize this model in
an experiment \cite{you05,You03,You032,Blais2, Mooji, Takayanagi, Deppe,verdu} (section IX).

\begin{figure}[]
\includegraphics[width=\columnwidth]{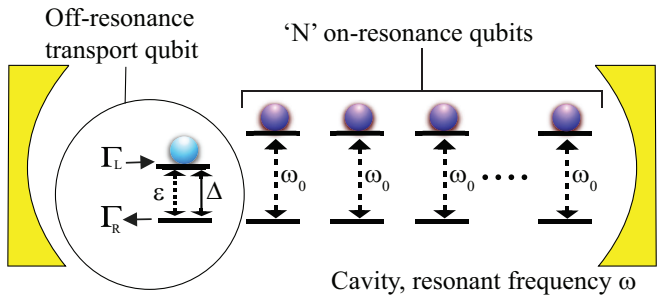}
\vspace{0.0cm} \caption{(Color Online)  Geometry of the proposed
device.
  $N$ qubits are placed at the anti-node(s) of a `cavity' with resonant frequency $\omega$ (depending on specific realization, they might alternatively be placed at one anti-node, or at subsequent anti-nodes, to minimize qubit-qubit interactions).  Their energy splittings are on-resonance with the
 oscillator $\omega_0=\omega$.
        An additional `open' transport qubit (TQ, shown in the left) is also coupled to the cavity off-resonance ($\epsilon \neq \omega$), and used to passively read-out the state of the cavity mode.  In the figure, the solid lines represent tunneling, while the dashed lines represent energy gaps.
        The properties of the transport qubit are defined by an
        energy splitting $\epsilon$, coherent tunneling rate
        $\Delta$ and transport rates $\Gamma_L$ and $\Gamma_R$.  In
        addition, the cavity mode has a cavity decay rate $\gamma_b$,
        not shown in the figure.
        Superconducting artificial atoms coupled to an on-chip cavity (e.g. a quantized LC oscillator or microwave transmission line) are a feasible
    realization using current technology.  The transport qubit can be realised using a charge qubit in the transport regime, i.e., a superconducting single electron transistor.
    Alternatively, a large number of qubits in the form of two-level atoms in an atom chip, coupled to a transmission line, has recently been proposed as a way to
    realise the large-$N$ Dicke model \cite{verdu}.
      }\label{RWA1}
\end{figure}

We point out that the properties we investigate here require the
precise control of the couplings between qubits and the resonator,
and access to a very strong coupling regime, both of which are
difficult to achieve. However it was recently shown~\cite{Goto2008}
that the generalized Dicke model, a variation of the Dicke model
where the couplings between the $N$ qubits and the cavity are
inhomogeneous, still has all the critical properties of the standard
Dicke model. This indicates the universality of the Dicke phase
transition, as well as making an experimental realization more
feasible. Furthermore, the critical point in the transition relies
on the relative values of the coupling strengths and the
level-splittings of the qubits. Thus, while the qubit-boson coupling
strengths are not tunable in a real experiment, the level splittings
of the qubits can typically be controlled by external parameters
(e.g., in the case of superconducting flux qubits via an external
magnetic flux \cite{Chen}).  Furthermore, a realization using Raman
transitions in atoms in an optical cavity has been proposed as a
method to reach a controllable strong coupling regime~\cite{Dimer}.

\section{Dicke Superradiant phase transition.}

Historically, the Dicke Hamiltonian (DH) describes the dipole
interaction between $N$ ``atoms'' and $n_b$ bosonic field modes.
Typically \cite{Wang73} the atoms are considered to be at fixed
sites within a cavity of volume $V$.  The atoms are assumed to be
well separated, and thus non-interacting.  Hereafter we refer to the
atoms as `qubits', and the additional measurement qubit as the
`transport qubit'.
 To observe critical phenomena we consider
the single-mode case with $n_b=1$.  
  We do not
make the rotating wave approximation, allowing the model to describe
both weak and strong coupling regimes (and we omit the 
$\vec{A}^2$ term).

Previous work \cite{Emary02,Lambert04} on this model has shown an
exact analytical solution in the limit $N\rightarrow \infty$.
Furthermore, the transition was characterized as a breaking of
parity symmetry at a particular value of the coupling between qubits
and cavity (denoted by $\lambda$, with the critical value being
$\lambda_c$). Both the qubits and the cavity bosonic degrees of
freedom become `macroscopically occupied' (i.e., of $O(N)$, the
number of qubits) in the regime above the critical point $\lambda >
\lambda_c$.  For finite arrays of qubits, $N$, the system is known
to exhibit power-law scaling \cite{vidal}, quantum
chaos~\cite{Emary02}, and critical entanglement
\cite{Lambert042,Lambert04}.

Several proposals for an experimental realisation of this system
have already been made.  For example, Dimer et al \cite{Dimer}
proposed a cavity QED realisation, and discussed in detail the
effect of the cavity decay on the phase transition.   In another
work, Chen et al \cite{Chen} proposed using superconducting charge
qubits coupled to an optical cavity, so that the critical properties
can be observed in the optical mode using heterodyne detection. In
addition they proposed observing the phase transition as a function
of level splitting, as discussed earlier.

\section{Dicke Hamiltonian.}
 The single-mode Dicke Hamiltonian
is defined as \beq H_D &=&
   \omega_0 \sum_{i=1}^N s_z^{(i)}
    + \omega  a^\dagger a
    + \sum_{i=1}^N
        \frac{\lambda}{\sqrt{N}} (a^\dagger + a)\
        (s^{(i)}_+ + s^{(i)}_-)
  \nonumber \\
  &=&
  \omega_0 J_z + \omega a^\dagger a
  + \frac{\lambda}{\sqrt{N}} (a^\dagger + a)(J_+ + J_-),
\label{DHam1} \eeq where $J_z=\sum_{i=1}^N s_z^{i}$,
$J_{\pm}=\sum_{i=1}^N s_{\pm}^{i}$ are collective angular momentum
operators for a pseudo-spin of length $j=N/2$.  These operators obey
the usual angular momentum commutation relations,
$[J_z,J_{\pm}]=\pm J_{\pm}$ and $[J_+,J_-]= 2J_z$. 
The frequency $\omega_0$ describes the qubit level splitting,
$\omega$ is the oscillator field frequency, and $\lambda$ the
qubit-field coupling strength. Because of their mutual interaction
with the oscillator field the qubits are not independent.  The
$\lambda/\sqrt{N}$ scaling is important to realise the thermodynamic
limit.   It essentially bosonises the low-energy part of the state
space of the collective angular momentum.  Physically, this scaling
implies that the density of qubits is constant, so that the cavity
volume becomes larger as $N$ is increased, consequently reducing the
electric field density and thus the effective interaction with each
individual qubit.

First, we will show analytical results for an entirely passive
measurement of the system, using an off-resonance ancillary qubit
with current transport.  Secondly, we will treat the back-action of
the ancillary qubit as a fully quantum interaction, with Markovian
transport properties, and including decay terms for the cavity. We
will see how this alters the final current measurements, as well as
how it changes the properties of the phase transition.

\section{Master Equation.}

To take into account the back-action of the transport qubit on the
Dicke Hamiltonian, we can model the whole system using a master
equation, \beq
\frac{d}{dt}\rho(t) &=& L[\rho(t)]= -i[H,\rho(t)] +L_0[\rho(t)]\\
H&=& H_D + H_{TQ}+H_{\mathrm{int}}, \quad L_0=L_{TQ} + L_{C}
\nonumber \eeq where \beq H_{TQ}=\epsilon \sigma_z + \Delta
\sigma_x,\eeq is the Hamiltonian of the transport qubit, \beq
H_{\mathrm{int}}=g \sigma_z a^{\dagger}a,\eeq where $
H_{TQ}=\epsilon \sigma_z + \Delta \sigma_x,$ is the Hamiltonian of
the {\em transport qubit (TQ)}.  Here $\epsilon$ is the level
splitting, and $\Delta$ the coherent tunneling within the TQ.
 $ H_{\textrm{int}}=g \sigma_z a^{\dagger}a,$ is the off-resonance
dispersive interaction between Dicke system and TQ, $L_{TQ}$
contains the transport properties of the TQ \cite{Lambert082}, and
$L_C$ contains cavity damping terms (e.g., photons leaking from the
cavity).
\begin{eqnarray}
L_{TQ}[\rho(t)]&=&-\frac{\Gamma_L}{2}\left[s_L s_L^\dagger
\rho(t) -
2s_L^\dagger \rho(t)s_L + \rho(t)s_L s_L^\dagger\right]\nonumber\\
&-&\frac{\Gamma_R}{2}\left[s_R^\dagger s_R \rho(t) - 2s_R
\rho(t)s_R^\dagger + \rho(t)s_R^\dagger s_R\right]\\
L_C&=&-\frac{\gamma_b}{2}\left[  a^\dagger a \rho-2a \rho a^\dagger
+ \rho a^\dagger a\right]
\end{eqnarray}
where \beq s_L=\ket{0}\bra{L},\,\,\,\,\,\,\,\,\,\, s_L^{\dagger}=\ket{L}\bra{0},\\
 s_R=\ket{0}\bra{R},\,\,\,\,\,\,\,\,\,\, s_R^{\dagger}=\ket{R}\bra{0},
 \eeq
$\Gamma_L$ and $\Gamma_R$ are the left/right tunneling rates for the
TQ, and $\gamma_b$ is the decay rate of photons out of the cavity
(throughout, we set $\hbar=1$).  Here $\rho(t)$ is the density
matrix describing the state of the qubit-array, cavity, and
transport qubit system.

\section{Passive Measurement.}

If we assume no back-action from the transport qubit onto the Dicke
model, the problem is very simple.  However, the form of the
interaction between the transport qubit and the effective cavity is
still important. As mentioned, off-resonance, $\epsilon \ll \omega$,
we assume the interaction is dispersive \cite{Clerk},
$H_{\mathrm{int}}= g\sigma_z a^{\dagger}a.$   For an entirely
non-destructive passive measurement (with no feedback), the state of
the ancillary transport qubit is then just shifted by the occupation
of the transmission line (i.e. considering the mean-field of
Eq.~[3]) , \beq H_{TQ} \approx (\epsilon +
g\ex{a^{\dagger}a})\sigma_z.\eeq

We are able to calculate the analytical values of
$\ex{a^{\dagger}a}$ in the limit $N\rightarrow \infty$.  The
transport properties are easily calculated using a
counting-statistics approach, which has been well summarised
elsewhere \cite{Lambert082}. Thus, the current and zero-frequency
current-noise measured through the ancillary qubit is simply given
by, \beq \frac{I}{e}= \frac{T_c^2 \Gamma_R}{T_c^2
(2+\Gamma_L/\Gamma_R) + \Gamma_R^2/4 + (\epsilon +
g\ex{a^{\dagger}a})^2}, \eeq
\begin{widetext} \beq S(0)=2eI\left[1- 8\Gamma_L T_c^2 \frac{4
(\epsilon+g\ex{a^{\dagger}a})^2 (\Gamma_R - \Gamma_L) +
\Gamma_R(3\Gamma_L\Gamma_R + \Gamma_R^2 + 8T_c^2)}{[4T_c^2(2\Gamma_L
+ \Gamma_R) + \Gamma_L\Gamma_R^2 + 4(\epsilon +
g\ex{a^{\dagger}a})^2\Gamma_L]^2}\right]. \eeq
\end{widetext}
In the limit $N\rightarrow \infty$ the Dicke Hamiltonian has two
distinct solutions, corresponding to the two phases of the
transition.  In the superradiant phase both cavity and qubit array
have a macroscopic mean field displacement.

In the lower, `normal phase', we define the occupation of the cavity
$\ex{a^{\dagger}a}$ by an effective temperature $T$ and frequency
$\Omega$, \beq \ex{a^{\dagger}a} = \left(\frac{m \Omega}{4\omega} +
\frac{\omega}{4m\Omega}\right)\coth\left(\frac{\Omega}{2T}\right)-\frac{1}{2}.
\eeq Where $\Omega$ and $T$ depend on the eigenenergies of $H$: \beq
  [\epsilon_{\pm}^{(1)}]^2 &=& \frac{1}{2} \rb{\omega^2 + \omega_0^2 \pm
  \sqrt{(\omega_0^2 - \omega^2)^2 +16\lambda^2\omega \omega_0}}
\eeq where $\epsilon_-$ is only real for $\lambda \leq \lambda_c$,
giving the range of this solution.  The dependence of $T$ and
$\Omega$ on the eigenvalues is via the relations,\beq \label{effT}
  \cosh \beta \Omega &=& \left[1 + \frac{2\epsilon_-
  \epsilon_+}{(\epsilon_- - \epsilon_+)^2 c^2 s^2}\right],
   \\
  m \Omega &=& \left[\rb{1+\frac{2\epsilon_-\epsilon_+}{(\epsilon_-
  - \epsilon+)^2 c^2 s^2}}^2 - 1\right]^{1/2} \\
  &\times& \left[\frac{(\epsilon_- - \epsilon_+)^2 c^2
  s^2}{2(\epsilon_-s^2 + \epsilon_+ c^2)}\right],\\
  c&\equiv&
\cos \gamma^{(1)}, \quad s\equiv\sin \gamma^{(1)}, \\
\tan(2\gamma^{(1)}) &=& \frac{4 \lambda \sqrt{\omega
  \omega_0}}{(\omega_0^2 - \omega^2)}
\eeq where $\beta = 1/k_B T$.  These define two equations linking
the three parameters of the cavity/qubit system $\omega$,
$\omega_0$, $\lambda$,  and the three effective parameters of a
thermal oscillator $\beta$, $\Omega$, $m$. By setting one energy
scale of the original system such that $\omega=1$, and that of the
thermal oscillator such that $m=1$, we can uniquely define the
correspondence between the two systems.  We use the relations, \beq
\cosh(\beta \Omega)&=&1+2\epsilon_-\epsilon_+/D,\\
D&\equiv& [sc(\epsilon_- - \epsilon_+)]^2,\\
2\Omega/\sinh(\beta\Omega) &=& D/(\epsilon_- s^2 + \epsilon_+
c^2),\\
\Omega \sinh(\beta \Omega)&=&\frac{2\epsilon_- \epsilon_+ (1+
\epsilon_-\epsilon_+/D)}{(\epsilon_-s^2 +\epsilon_+c^2)},\\  \coth
(\beta \Omega/2)&=&[\cosh(\beta \Omega) +1]/\sinh(\beta \omega) \eeq
to obtain, \beq \ex{a^{\dagger}a}=\frac{(\epsilon_-s^2 +
\epsilon_+c^2)}{4}\left(\frac{m}{\omega} + \frac{\omega}{m\epsilon_-
\epsilon_+}\right). \eeq Thus, in this passive measurement regime,
in the large $N$ limit, the occupation of the bosonic mode (which is
an order parameter of the phase transition) diverges as $\epsilon_-
\rightarrow 0$ when $\lambda \rightarrow \lambda_c$.   In the next
section we discuss the effect of this on the current-measurement.

\section{Power-law scaling in transport properties.}
\subsection{Results}

We plot the current and current-noise in Figs.~\ref{igraph} and
\ref{sgraph}.   We immediately see that, at the critical point
$\lambda_c$, the large occupation of the cavity mode (which is
proportional to the number of qubits $N$) acts to blockade the
current flow (by ``pushing apart'' the internal energy levels of the
transport qubit). Similarly, the zero-frequency noise becomes
strictly Poissonian at the critical point.  This is a consequence of
the slow current and charge-dominated dynamics. Thus, both the
current and current-noise are operating as {\em signatures, or order
parameters, of the phase transition}, because of their direct
dependence on $\ex{a^{\dagger}a}$.

As mentioned earlier, in previous work \cite{Emary02, HL73} the
phase transition was studied as a function of multi-qubit-oscillator
coupling $\lambda$. However, the transition can also be observed for
a given constant $\lambda$, by tuning the energy level of the qubits
$\omega_0$. This is a more realistic approach with superconducting
qubits as a possible realisation. Qualitatively, the properties of
the transition are the same.  For instance, for $\lambda=0.1\omega$
the transition occurs when $\omega_{0,c}\rightarrow 0.04\omega$. The
sub-radiant phase occurs for $\omega_0 >\omega_{0,c}$, while the
super-radiant phase appears when $\omega_0 < \omega_{0,c}$, both of
which are experimentally-accessible regimes.  However, because the
interaction is off-resonance, the convergence to the correct scaling
behaviour requires much larger $N$.

\subsection{Scaling with the number ($N$) of qubits}

\begin{figure}[]
\includegraphics[width=\columnwidth]{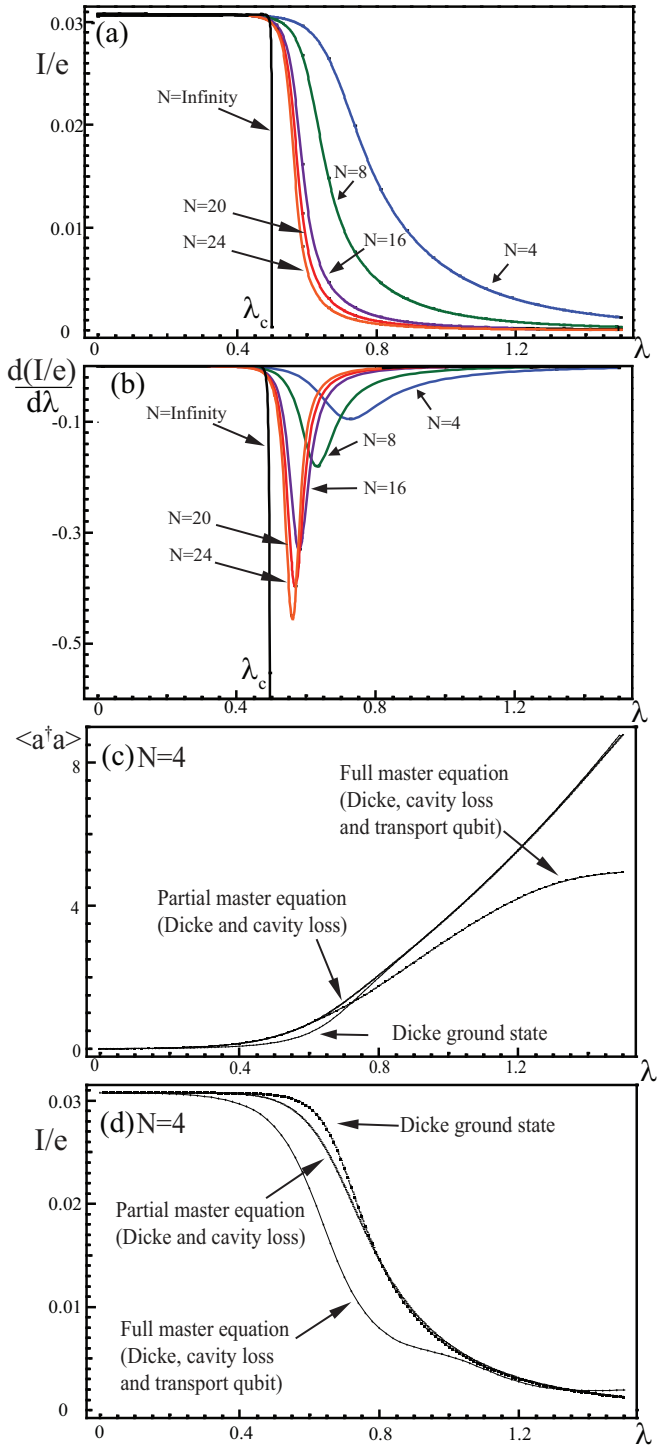}
\vspace{0.0cm} \caption{(Color online) (a) The current $I/e$ versus
multi-qubit-oscillator coupling $\lambda$ through the transport
qubit for $T_c=0.1$, $\Gamma_L=\Gamma_R=0.1$, $\epsilon=0$,
$\omega=\omega_b=1$, $g=0.1$ for $N=4, 8, 16, 20, 24, \infty$.  (b)
The derivative of the current through the transport qubit for the
same parameter set, versus $\lambda$.  Figures (c) and (d) show one
particular data curve ($N=4$, $\gamma_b=0.1$) for the bosonic
occupancy $\ex{a^{\dagger}a}$ and the current $I/e$ for the three
different approximations; zero back-action (ground state of the pure
Dicke model), master equation with cavity damping, and master
equation with cavity damping and transport qubit feedback.
}\label{igraph}
\end{figure}

\begin{figure}[]
\includegraphics[width=\columnwidth]{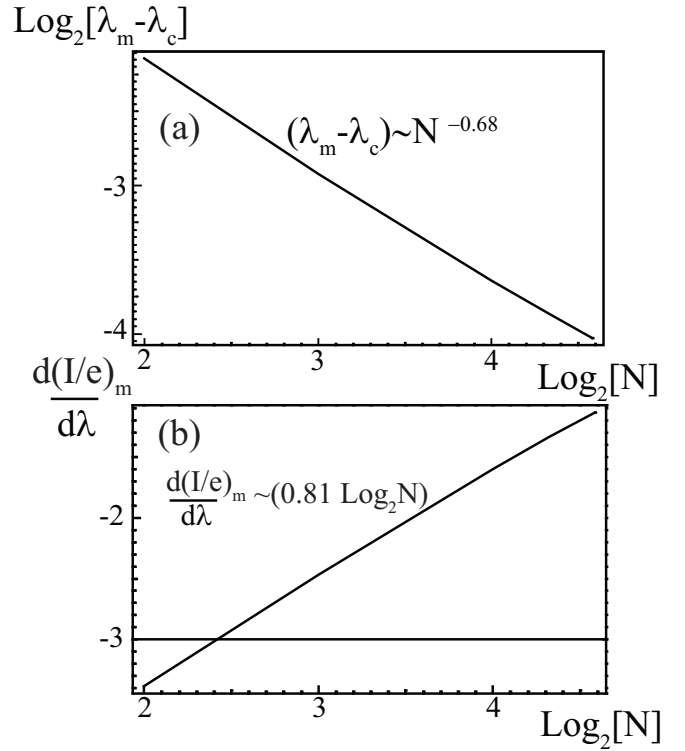}
\vspace{0.0cm} \caption{(Color online) (a) shows the scaling with
$N$ and scaling exponent of the position ($\lambda_m$) of the
minimum of the current derivative: $(\lambda_{m} - \lambda_c)
\propto N^{-0.68\pm 0.05}$.  Figure (b) shows the scaling of the
value of the current at this minimum point to be $\left(\frac{d
(I/e)}{d \lambda}\right)_m \propto (0.81 \pm 0.05) \log_2 N$. The
parameters used here are $T_c=0.1$, $\Gamma_L=\Gamma_R=0.1$,
$\epsilon=0$, $\omega=\omega_b=1$, $g=0.1$ with data taken at $N=4,
8, 16, 20, 24,40,60$.}\label{scalegraph}
\end{figure}

\begin{figure}[]
\includegraphics[width=\columnwidth]{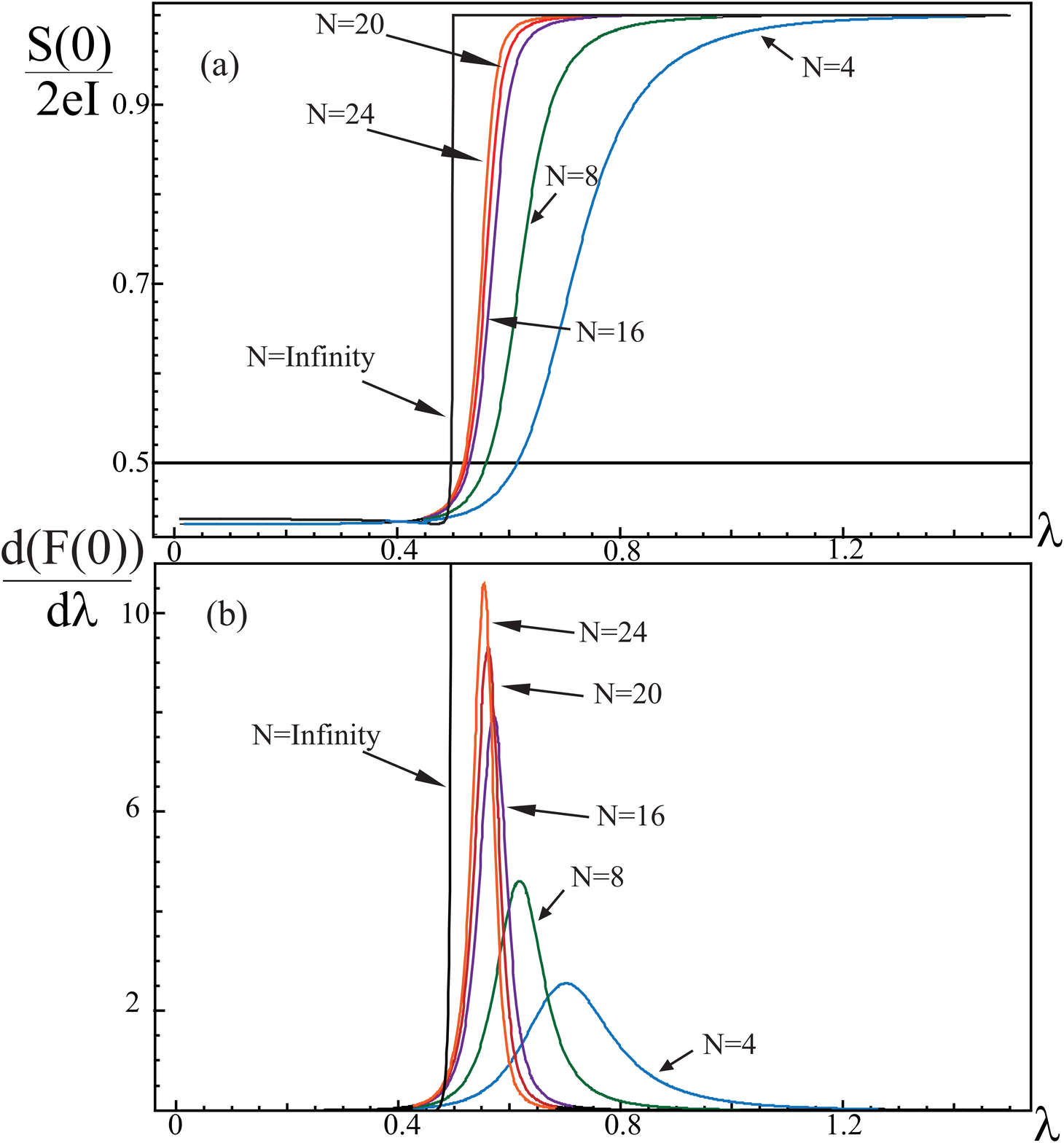}
\vspace{0.0cm} \caption{(Color online) (a) The current-noise
$F(0)=S(0)/2eI$ versus multi-qubit-oscillator coupling $\lambda$
for: $T_c=0.1$, $\Gamma_L=\Gamma_R=0.01$, $\epsilon=0$,
$\omega=\omega_b=1$, $g=0.1$ and for $N=4, 8, 16,20,24, \infty$. (b)
 The derivative $\frac{d F(0)}{d \lambda}$ versus $\lambda$.   The peak scales as a power law of $N$, similar to the minimum of the current derivative.}\label{sgraph}
\end{figure}

To observe power-law scaling with $N$, we must look at the
derivative of both the current and current-noise with respect to the
Dicke multi-qubit-oscillator coupling $\lambda$. The minimum value
of these derivatives will act as a signature of ``precursor
behaviour'', and from them we can extract the power-law dependence.
In Fig.~\ref{igraph}(b) we show the derivative of the current, and
in Fig.~\ref{scalegraph}(a) we see that the position of the minimum
of the current derivative scales as a power law in $N$ via \beq
(\lambda_m -\lambda_c )\propto N^{-0.68 \pm 0.05}.\eeq This matches
a previous result for the scaling of the entanglement entropy
\cite{Lambert04}. Similarly the value of the current at this minimum
point scales logarithmically as \beq \frac{d (I/e)}{d \lambda}_m
\propto (0.81 \pm 0.05) \log_2 N,\eeq as shown in
Fig.~\ref{scalegraph}(b).  The value of the current derivative obeys
similar scaling laws.

Vidal et al \cite{vidal} studied the scaling, in $N$, at the
critical point $\lambda_c$ of several properties of the Dicke model.
They predicted a scaling exponent for $1/\ex{a^{\dagger}a}^2$ of
$\alpha = 2/3$.  These exponents are different from those we observe
here, as they describe behaviour of quantities measured exactly at
$\lambda_c$.  To extract the same exponents from our numerics would
require very large values of $N$.  However a recent numerical study
by Chen et al \cite{chendicke08} describes a scheme where such
exponents can be calculated efficiently for large $N$, and
confirmed~\cite{vidal} the correct exponents for some of these
quantities.

\section{Back-action and cavity loss.}

To take into account both the back-action of the transport qubit,
and the loss of photons from the cavity due to coupling to the
environment, we must solve the entire master equation numerically.
This is a non-trivial task, even with state-of-the-art numerics, and
requires careful use of sparse-matrix techniques to increase
efficiency.

Dimer et al \cite{Dimer} investigated the thermodynamic limit of the
Dicke model including losses from the bosonic cavity. They found
that the critical point was shifted from its normal position as a
function of the cavity loss $\gamma_b$.  In Figs. 2(c) and 2(d) we
do the same for the finite-$N$ case, comparing the three possible
regimes: zero back-action and no cavity loss,  zero back-action with
cavity loss, and a full treatment of cavity loss and back-action.

In figure 2(c) we see that around the critical point the occupancy
of the bosonic cavity is almost exactly the same for both master
equation treatments, but differs slightly from the ground state
Dicke case. Furthermore the strong coupling limit for the full
master equation treatment saturates because of the bosonic Hilbert
space cut-off needed in solving this complex problem. Furthermore,
in figure 2(d) we see that the full treatment of the combined
transport qubit/Dicke model shows a reduced current profile compared
to the two situations with zero back-action.   This is also the case
for other values of $N$.

However the coupling to the qubit, and the loss of energy from the
cavity, has less obvious effects
  on the properties of the phase transition itself.  In particular,
  the parity,
  \beq
  \Pi=\exp[i \pi (a^{\dagger}a + J_z + j)]
  \eeq
  is no longer conserved, and the steady state will contain
  components of both the ground state and excited states of $H_D$.   Because of this, and the restrictions
  on the number of spins we can efficiently model, it is not possible to extract exponents from this data.
    However, we expect the large-$N$ limit to still exhibit features of the phase transition, as predicted by Dimer et
al \cite{Dimer}.

\section{Signatures of Quantum Chaos.}

Quantum chaos is a characteristic of non-integrable quantum systems.
Emary et al \cite{Emary02} extensively studied the (closed) Dicke
model and its chaotic properties.   In the finite-$N$ regime they
showed that the eigenvalue spectrum of the Dicke model fitted that
of the Wigner-Dyson distribution~\cite{GUHR} when the qubit-boson
coupling was around the critical point $\lambda\approx \lambda_c$.
Thus, the chaotic behaviour is understood to be a `precursor' of the
phase transition, driven by the parity conservation at the critical
point.

Here we extend their work by identifying similar distributions in
the eigenvalues, \beq \chi_i = i(E^L_i) + \nu_i \eeq of the {\em
Liouvillian} $L$ which include imaginary components $i(E^L_i)$ from
$H_D$, as well as real components $\nu_i$ from the cavity loss
terms. Here we ignore the back-action and electron transport terms
here, and focus on the effect of cavity damping on the level
statistics.

For the pure-state case (no cavity losses), the von Neumann equation
of motion, \beq \frac{\rho(t)}{dt} = -i[H,\rho(t)] \eeq can be
written as a set of $N_H^2$ coupled equations of the matrix elements
of $\rho$, where $N_H$ is the dimension of the Hilbert space for the
system described by the Hamiltonian $H$.  If $H$ has $N_H$
eigenvalues $E_k$, $k=1,...,N_H$, and we take matrix elements
according to the eigenbasis of $H$, then we can write these linear
equations as a diagonal matrix with $N_H^2$ imaginary eigenvalues
\beq E^L_{i=j\times k}=\sum_{k,j=1}^{N_H}(E_{k} - E_{j}).\eeq Every
possible energy gap (not just nearest neighbor) in the spectrum of
$H$ has an eigenvalue in $L$.

In Fig.~(4) we show the positive branch of the imaginary components
of the eigenvalues of $L$ for $N=6$, $\lambda = \lambda_c$, after
removal of the $N_H$ zeros, i.e. the stationary states, and the
probability distributions of these components. Even though it is not
possible to unfold this spectrum, and all possible level spacings
are present, still we see some characteristics of the `picket-fence'
distribution~\cite{kus} of the Rabi Hamiltonian and the universal
Wigner-Dyson distribution~\cite{GUHR}. We point out that the
eigenvalues of this matrix, which is a particular representation of
the superoperator $L$, determine many of the higher-order transport
properties, like the frequency dependant noise.  This is also seen
in scattering theory~\cite{beenakker}. Further analytical work needs
to be done to make a strong connection between measurable transport
quantities, random matrix theory, and quantum chaos.

\begin{figure}[]
\includegraphics[width=\columnwidth]{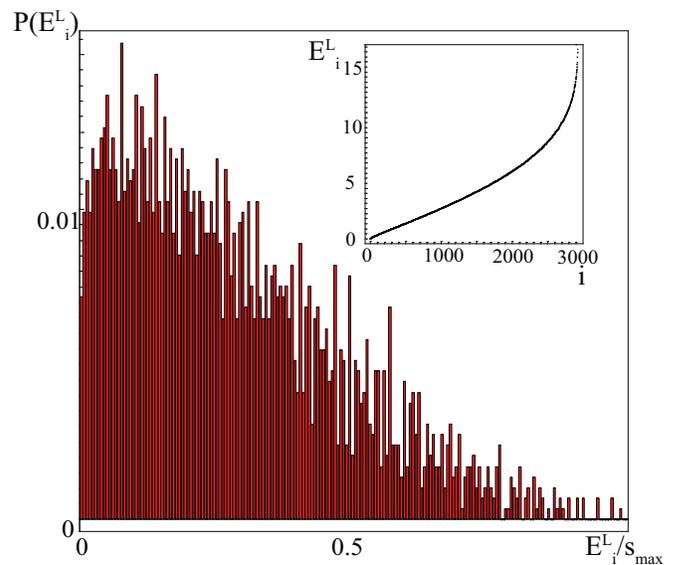}
\vspace{0.0cm} \caption{(Color online) The inset shows an example of
the positive imaginary components of the eigenvalues ($E^L_i$) of
the Liouvillian for the damped Dicke model and the main figure shows
their probability distribution $P(E^L_i)$, normalised to the maximum
energy gap $S_{\mathrm{max}}$, for $N=6$, $\lambda=\lambda_c$, and
$\gamma_b=0.1$. While this contains every possible eigenvalue
seperation of the Hamiltonian (up to  numerical bosonic cut-off),
and has not been unfolded to remove secular variations, level
repulsion is still visible. }\label{cgraph}
\end{figure}

\section{From circuit QED to the Dicke Model.}

Al-saidi and Stroud \cite{Stroud02} have studied a realization of
the Dicke model using Josephson Junctions coupled to an
electromagnetic cavity.  Operating in the regime `between' charge
and flux qubits they showed that, given the right parameters, the
higher-lying levels of each junction can be neglected.   In the same
way, it is possible to derive the Dicke Hamiltonian,
Eq.~(\ref{DHam1}), from the Hamiltonian describing superconducting
qubits interacting with a cavity.  The proposal and realization of
cavity QED~\cite{you05,You03,You032,Blais2, Mooji, Takayanagi,
Deppe} in a circuit was an important development for quantum optics
and condensed matter, and thus the observation of strong many-body
effects in these systems is a natural extension of previous work.

Alternatively, a large number of qubits, in the form of two-level
atoms in an atom-chip, coupled to a transmission line, was recently
proposed as a way to realise the large-$N$ Dicke model \cite{verdu}.

\section{Conclusions}

In conclusion, we have shown that current and current-noise
measurements could be used to test for criticality in an `on chip'
experiment. We extracted scaling exponents for the Dicke phase
transition from semi-analytical and numerical modelling, and
illustrated how quantum chaos, a precursor behaviour to the phase
transition, is retained in an open-system environment.

\acknowledgements

We thank S. Ashhab and I. Mahboob for helpful discussions.  This
work is supported partially by the National Science Council, Taiwan
under the grant number 95-2112-M-006-031-MY3. FN acknowledges
partial support from the National Security Agency (NSA), Laboratory
for Physical Sciences (LPS), Army Research Office (ARO), National
Science Foundation (NSF) Grant No. EIA-0130383, JSPS-RFBR contract
No. 06-02-91200, and CTC program supported by the Japan Society for
Promotion of Science (JSPS).

\bibliography{bibliography}

\end{document}